\documentclass[cits]{PoS}

\usepackage[utf8x]{inputenc}
\usepackage{mciteplus}

\newcommand{\der}{\mathrm{d}}
\newcommand{\rt}{{\mathbf{r}_T}}

\newcommand{\bt}{{\mathbf{b}_T}}
\newcommand{\st}{{\mathbf{s}_T}}

\newcommand{\qt}{{\mathbf{q}_T}}
\newcommand{\kt}{{\mathbf{k}_T}}

\newcommand{\cf}{C_\mathrm{F}}

\newcommand{\qs}{Q_\mathrm{s}}

\newcommand{\lqcd}{\Lambda_{\mathrm{QCD}}}
\newcommand{\as}{\alpha_{\mathrm{s}}}

\title{Dipole amplitude with uncertainty estimate from HERA
data and applications in Color Glass Condensate
phenomenology}

\ShortTitle{Dipole model fits and applications}

\author{T. Lappi\\
Department of Physics, University of Jyväskylä, %
 P.O. Box 35, 40014 University of Jyv\"askyl\"a, Finland\\
 and \\
 Helsinki Institute of Physics, P.O. Box 64, 00014 University of Helsinki,
Finland \\
        E-mail: \email{tuomas.v.v.lappi@jyu.fi}}

\author{\speaker{H. Mäntysaari}\\
        Department of Physics, University of Jyväskylä, %
 P.O. Box 35, 40014 University of Jyv\"askyl\"a, Finland\\
        E-mail: \email{heikki.mantysaari@jyu.fi}}

\abstract{We determine the initial condition for the small-$x$ evolution equation (BK) 
from the HERA deep inelastic scattering data using a new parametrization that also keeps the unintegrated gluon distribution positive. The obtained dipole amplitude and its uncertainty estimate can be used to compute single inclusive particle production in proton-proton and proton-nucleus collisions. We argue that one has to use consistently the proton transverse area measured in DIS and the total inelastic cross section when calculating the single inclusive cross section. This leads to a midrapidity nuclear modification factor $R_{pA}$ that approaches unity at large transverse momentum, independently of the center-of-mass energy. }

\FullConference{XXII. International Workshop on Deep-Inelastic Scattering and Related Subjects,\\
		28 April - 2 May 2014\\
		Warsaw, Poland}

\begin{document}

\section{Introduction}
The Color Glass Condensate is an effective field theory that provides a convenient framework to describe strongly interacting systems at high energy where non-linear phenomena, such as gluon recombination, become important. These nonlinearities are further increased when the target is changed from a proton to a heavy nucleus, due to the $A^{1/3}$ scaling of the gluon densities.

A key ingredient in the CGC calculations is the dipole-proton amplitude, whose evolution in Bjorken-$x$ (or equivalently, energy) is given by the the BK equation~\cite{Balitsky:1995ub,Kovchegov:1999yj} (with running coupling corrections derived in Ref.~\cite{Balitsky:2006wa}). Perturbative techniques can be used to derive the BK equation, but its initial condition, the dipole-proton amplitude at initial $x$, is a non-perturbative input. It can be obtained by performing a fit to proton structure function data measured in deep inelastic scattering (DIS) experiments.

A crucial test for the CGC framework comes from the fits to the combined proton structure function data from the H1 and ZEUS experiments at HERA~\cite{Aaron:2009aa,Abramowicz:1900rp}. 
A good fit to this precise data can be obtained by using a simple parametrization for the initial dipole amplitude, see Refs.~\cite{Lappi:2013zma,Albacete:2010sy}.

In this work we discuss how the dipole amplitude is obtained from the HERA DIS data and how the obtained  dipole-proton amplitude can be used to describe proton-proton and proton-nucleus collisions. This is reported in more detail in Ref.~\cite{Lappi:2013zma}. We also report ongoing work on the analysis of how tightly the DIS data constrains the initial conditions: we evaluate the uncertainty estimate for the dipole amplitude and study the propagation of these uncertainties to other observables.

\section{Fitting the dipole amplitude}
The H1 and ZEUS experiments have measured the proton structure functions $F_2$ and $F_L$, and published the precise combined results for the reduced cross section $\sigma_r$~\cite{Aaron:2009aa}, which is a function of the proton structure functions:
\begin{equation}
	\sigma_r(y,x,Q^2) = F_2(x,Q^2)-\frac{y^2}{1+(1-y)^2} F_L(x,Q^2),
\end{equation}
where $x$ is the Bjoken-$x$, $Q^2$ is the virtuality of the photon and $y$ stands for the inelasticity. The structure functions can be computed from the Color Glass Condensate framework by evaluating the virtual photon-proton cross section
\begin{equation}
		\sigma_{T,L}^{\gamma^*p}(x,Q^2) = 2\sum_f \int \der z \int \der^2 \bt \der^2 \rt |\Psi_{T,L}^{\gamma^* \to f\bar f}(\rt,z)|^2 N(\bt, \rt, x),
\end{equation}
where $\Psi_{T,L}^{\gamma^* \to f\bar f}$ is the photon light cone wave 
function describing how the photon fluctuates to a quark-antiquark pair,
computed from light cone QED~\cite{Kovchegov:2012mbw}. The QCD dynamics is encoded in the dipole-proton amplitude $N$, and the summation is over active quark favours. In this work we do not take into account the heavy quarks, but we note that if the heavy quarks are included one should also include in the analysis the measured charm contribution to $\sigma_r$ from Ref.~\cite{Abramowicz:1900rp}. We assume that the impact parameter dependence can be factorized and we replace $2\int \der^2 \bt$ by $\sigma_0$ (twise the proton DIS area) which is a fit parameter.

For the dipole amplitude at $x_0=0.01$ we use a modified McLerran-Venugopalan model~\cite{McLerran:1994ni}:
\begin{equation}
\label{eq:aamqs-n}
	N(\rt,x=x_0) = 1 - \exp \left[ -\frac{(\rt^2 Q_{s,0}^2)^\gamma}{4} \ln \left(\frac{1}{|\rt| \lqcd}+e_c \cdot e\right)\right],
\end{equation}
where we have generalized the AAMQS~\cite{Albacete:2010sy} form (labeled as MV$^\gamma$) by also
allowing the constant inside the logarithm, which plays a role of an infrared cutoff, to be different from $e$. The other fit parameters are the
anomalous dimension $\gamma$ and the initial saturation scale $Q_{s,0}^2$. The last fit parameter is the scale at which the running coupling is evaluated in transverse coordinate space, which we write as $\mu^2=4C^2/r^2$ and fit $C^2$. We fit three different initial conditions to the HERA $\sigma_r$ data, and the fit result is shown in Table~\ref{tab:params}.

The first initial condition used is the standard MV model where $\gamma=e_c=1$. The second parametrization considered here is the MV$^\gamma$ in which $\gamma$ is a fit parameter but $e_c=1$. The third parametrization is labeled as MV$^e$, and it has $\gamma=1$ but $e_c$ is free. A motivation for the last parametrization is that the Fourier-transform of $S(\rt)=1-N(\rt)$, which is proportional to the unintegrated gluon distribution, is not positive definite when the MV$^\gamma$ parametrization is used.

\section{Dipole amplitude for nuclear targets}
We generalize the dipole-proton amplitude to dipole-nucleus scattering by using the optical Glauber model, and write the dipole-nucleus amplitude $N^A$ as
\begin{equation}
	N^A(\rt,\bt) = 1-\exp \left( -\frac{AT_A(\bt)}{2} \sigma_\mathrm{dip}^p \right),
\end{equation}
where $\sigma_\mathrm{dip}^p$ is the total dipole-proton cross section. In order to satisfy the requirement $N^A\to 1$ at large dipoles we use a non-unitarized version of the dipole-proton cross section and obtain (for more details, see Ref.~\cite{Lappi:2013zma})
\begin{equation}
	N^A(\rt,\bt) = 1-\exp \left[ -A T_A(\bt) \frac{\sigma_0}{2} \frac{(\rt^2 Q_{s0}^2)^\gamma}{4} \left( \frac{1}{|\rt|\lqcd}+e_c\cdot e\right) \right].
\end{equation}

\section{Uncertainty analysis}
The experimental uncertainties can be propagated to the fitted dipole amplitude using the Hessian method~\cite{Pumplin:2001ct} where one uses a quadratic approximation
\begin{equation}
	\chi^2 \approx \chi_0^2 + \sum_{ij} H_{ij} (a_i - a_i^0)(a_j - a_j^0).
\end{equation}
Here the best fit parameters, that minimise $\chi^2$, are $S^0=\{a_i^0\}$, and the Hessian matrix $H_{ij}$ can be related to the second partial derivatives of $\chi^2$ with respect to the fit parameters. The eigenvectors of the Hessian matrix serve as an uncorrelated basis for the error sets of the fit parameters $S_i^\pm$. Using the error sets one can, following the procedure used in Ref.~\cite{Eskola:2009uj}, compute an uncertainty estimate for any quantity $X$ that depends on the dipole amplitude as
\begin{equation}
	(\Delta X^\pm)^2 = \sum_k \left[ \max \big/ \min \{ X(S_k^+)-X(S^0), X(S_k^-)-X(S^0),0\}\right]^2.
\end{equation}
In our preliminary analysis we construct very conservatively the error sets such that $\Delta \chi^2$, the difference of $\chi^2$ obtained by using the best fit set $S^0$ or the error set $S_k^\pm$, is chosen to be $\Delta \chi^2\approx 36$.

\section{Single inclusive particle production}
The gluon spectrum in heavy ion collisions can be obtained
by solving the classical Yang-Mills equations of
motion for the color fields. For $k_T\gtrsim \qs$ 
it has been shown numerically~\cite{Blaizot:2010kh}
that this solution is well approximated by the 
following $k_T$-factorized formula ~\cite{Kovchegov:2001sc}
\begin{equation}
\label{eq:ktfact-bdep}
\frac{\der \sigma}{\der y \der^2 \kt \der^2 \bt} = \frac{2 \as}{\cf \kt^2 } \int \der^2 \qt \der^2 \st \frac{\varphi_p(\qt,\st)}{\qt^2} 
	 \frac{\varphi_p(\kt-\qt,\bt-\st)}{(\kt-\qt)^2}.
\end{equation}
Here $\varphi_p$ is the dipole unintegrated gluon
distribution (UGD) of the 
proton~\cite{Kharzeev:2003wz,Blaizot:2004wu,Dominguez:2011wm} proportional to the two-dimensional Fourier transform of $1-N_A$, where $N_A$ is the dipole amplitude in the adjoint representation. The impact parameter dependece is assumed to factorize. For a detailed expressions, see Ref.~\cite{Lappi:2013zma}. This gives the invariant yield as
\begin{equation}
\label{eq:ktfact-pp}
	\frac{\der N}{\der y \der^2 \kt} = \frac{(\sigma_0/2)^2}{\sigma_\mathrm{inel}} \frac{\cf}{8\pi^4  \kt^2 \as} \int \frac{\der^2 \qt}{(2\pi)^2} \qt^2 \tilde S^p(\qt) 
		(\kt-\qt)^2 \tilde S^p(\kt-\qt),
\end{equation}
where $\tilde S^p$ is the two-dimensional Fourier transform of $1-N_A$.

Assuming that $|\kt|$ is much larger than the saturation scale of one of the
colliding objects we obtain the hybrid formalism result
\begin{equation}
	\label{eq:pp-hybrid}
	\frac{\der N}{\der y \der^2 \kt} = \frac{\sigma_0/2}{\sigma_\mathrm{inel}} \frac{1}{(2\pi)^2} xg(x,\kt^2) \tilde S^p(\kt),
\end{equation}
where $xg$ is the integrated gluon distribution function. For it we can use the conventional parton distribution function, and in this work the CTEQ LO~\cite{Pumplin:2002vw} pdf is used. Note that the overall normalization factor is obtained by using both the proton area measured in DIS, $\sigma_0/2$, and the inelastic proton-proton cross section $\sigma_\mathrm{inel}$, consistently in the calculation.

\begin{table}
\begin{center}
\begin{tabular}{|l||r|r|r|r|r|r|r|}
\hline
Model & $\chi^2/N$ &  
$Q_{s,0}^2$ [GeV$^2$] & $Q_s^2$ [GeV$^2$] & $\gamma$ & $C^2$ & $e_c$ & $\sigma_0/2$ [mb] \\
\hline\hline
MV & 2.76 & 0.104 & 0.139  & 1 & 14.5 & 1 & 18.81 \\
MV$^\gamma$ & 1.17 & 0.165 & 0.245 & 1.135 & 6.35 & 1 & 16.45 \\
MV$^e$ & 1.15 & 0.060 & 0.238  & 1 & 7.2 & 18.9 & 16.36 \\
\hline
\end{tabular}
\caption{Parameters from fits to HERA reduced cross section data at $x<10^{-2}$ and $Q^2<50\,\mathrm{GeV}^2$ for different initial conditions. Also the corresponding initial saturation scales $\qs^2$ defined via equation $N(r^2=2/\qs^2)=1-e^{-1/2}$ are shown. The parameters for the MV$^\gamma$ initial condition are obtained by the AAMQS collaboration \cite{Albacete:2010sy}.
}
\label{tab:params}
\end{center}
\end{table}

\section{Results}
\begin{figure}
\begin{minipage}[t]{0.48\linewidth}
\centering
\includegraphics[width=1.05\textwidth]{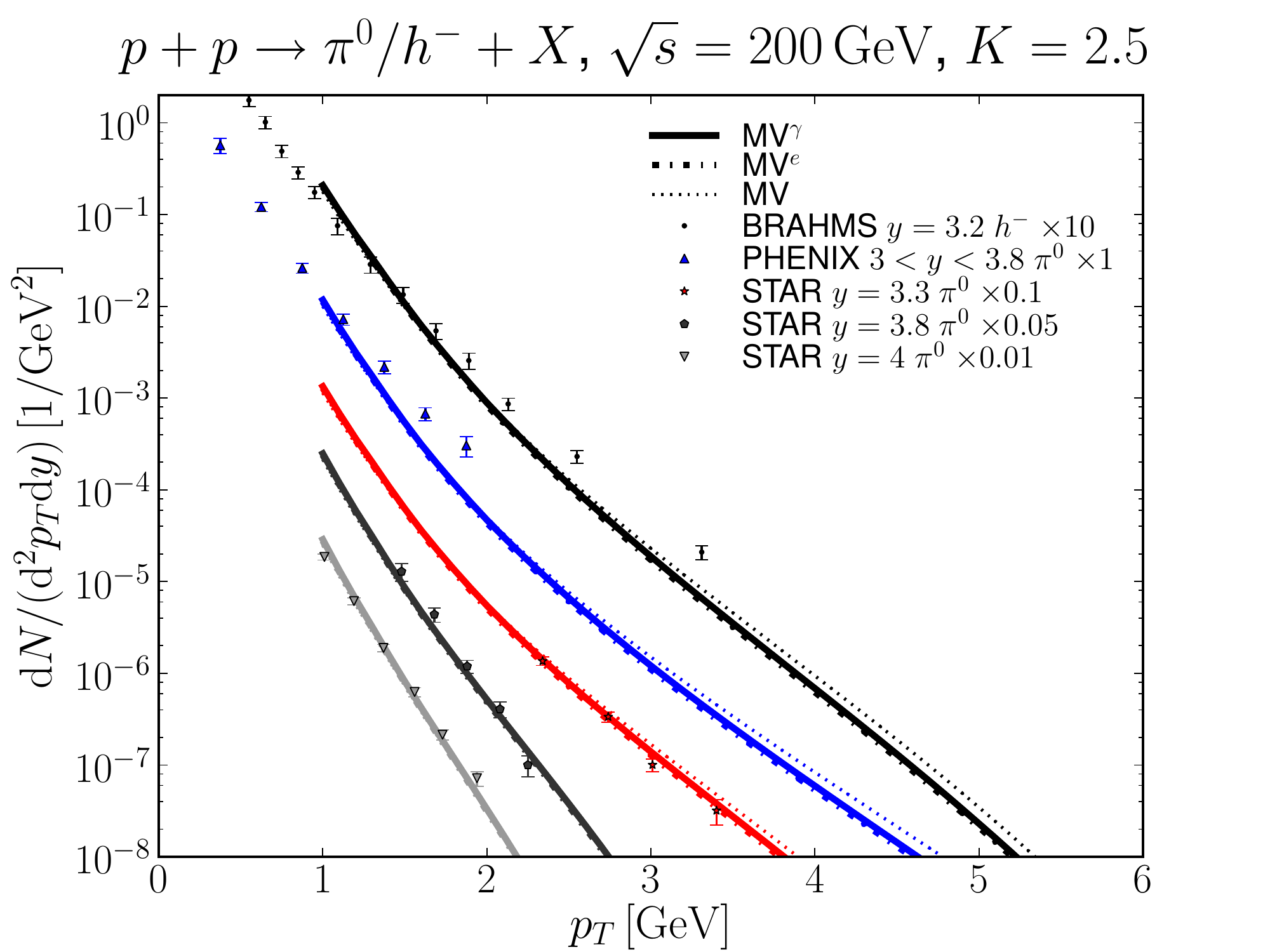} 
\caption{Single inclusive particle production at forward rapidities compared with the RHIC data~\cite{Adams:2006uz,Adare:2011sc,Arsene:2004ux}.}
\label{fig:rhic} 
\end{minipage}
\hspace{0.5cm}
\begin{minipage}[t]{0.48\linewidth}
\centering
\includegraphics[width=1.05\textwidth]{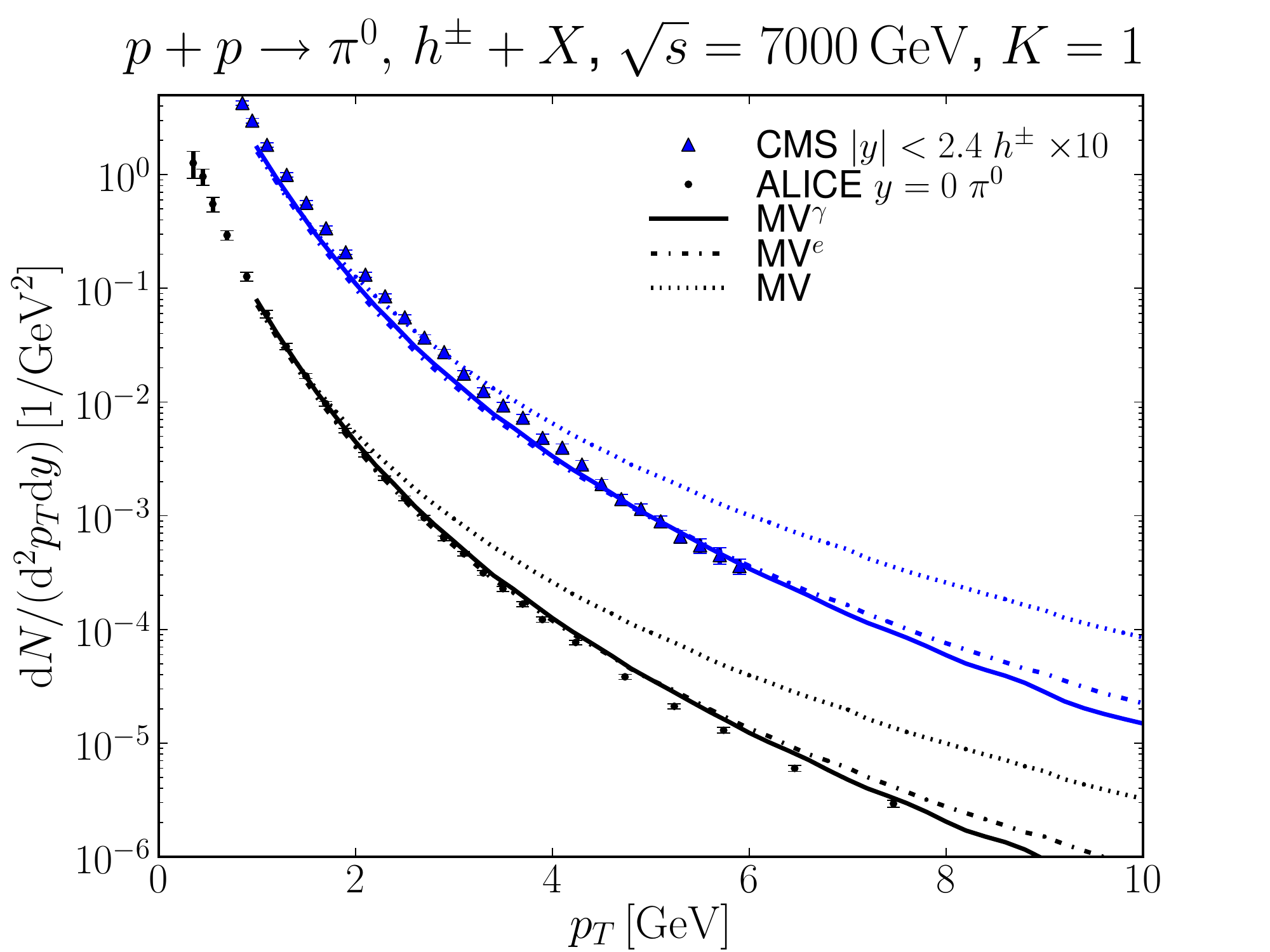} 
\caption{Charged hadron and $\pi^0$ production at midrapidity compared with the LHC data~\cite{Abelev:2012cn,Khachatryan:2010us}.}
\label{fig:lhc} 
\end{minipage}
\end{figure}

In Fig. \ref{fig:rhic} we show the single inclusive $\pi^0$ and negative hadron yields computed using the hybrid formalism and compared with the RHIC data~\cite{Adams:2006uz,Adare:2011sc,Arsene:2004ux}. As the calculation is done at leading order, it is not suprising that overall normalization does not agree with the data, but a normalization factor $K=2.5$ is needed. As we consider different proton areas consistently when deriving the hybrid formalism result and obtain a correct normalization factor for the LO calculation, the absolute value of the $K$ factor quantifies how much the LO result differs from the data. The comparison with the LHC data~\cite{Abelev:2012cn,Khachatryan:2010us} is shown in Fig. \ref{fig:lhc}. We notice that even though the standard MV model gives relatively good agreement with the $\sigma_r$ data, and especially works well with the RHIC forward data, comparison with the midrapidity LHC measurements clearly rules out the MV parametrization.

\begin{figure}
\begin{minipage}[t]{0.48\linewidth}
\centering
\includegraphics[width=1.05\textwidth]{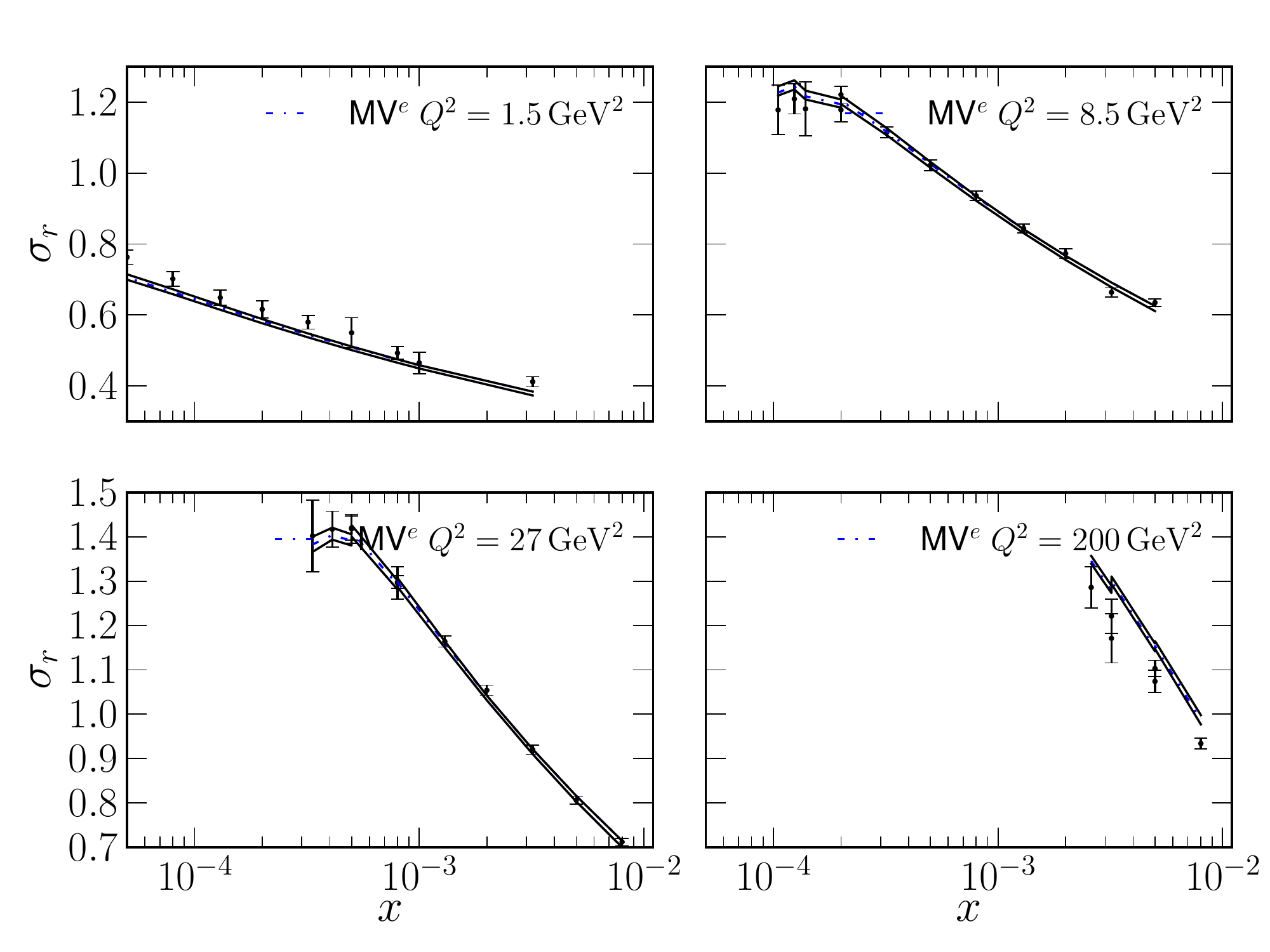} 
\caption{The reduced cross section measured at HERA~\cite{Aaron:2009aa} compared with the calculation done using the MV$^e$ parametrization and its error sets.}
\label{fig:hera} 
\end{minipage}
\hspace{0.5cm}
\begin{minipage}[t]{0.48\linewidth}
\centering
\includegraphics[width=1.05\textwidth]{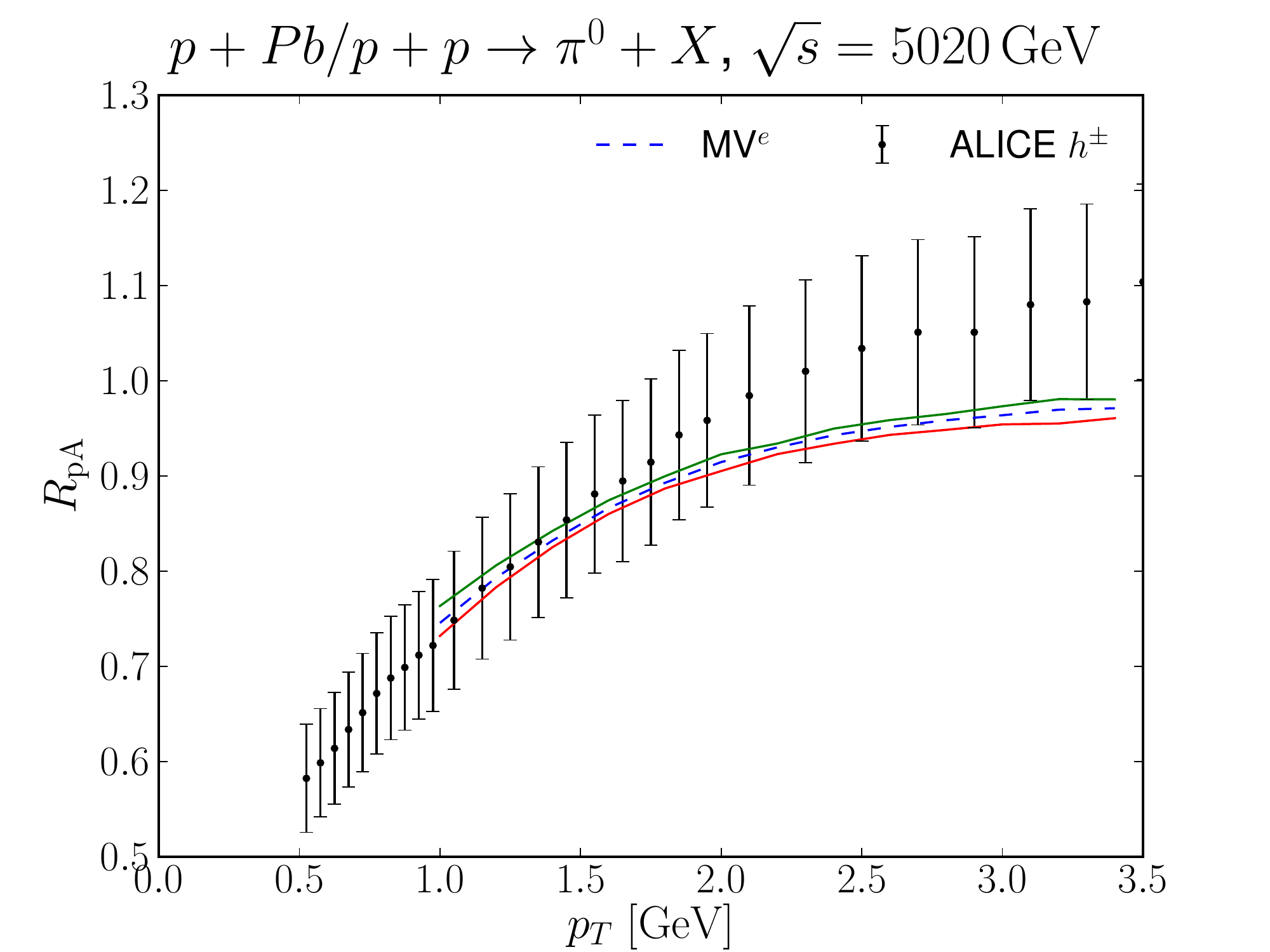} 
\caption{Nuclear suppression factor for $\pi^0$ computed using the MV$^e$ model and its error sets compared with the ALICE charged hadron $R_{pA}$~\cite{ALICE:2012mj}.}
\label{fig:rpa} 
\end{minipage}
\end{figure}

The computed reduced cross section and the uncertainty estimates are compared with the HERA data in Fig.~\ref{fig:hera}. Due to the accuracy of the data the uncertainty band is quite narrow, but the agreement with the data is very good except at largest $Q^2$ that is not included in the fit. As a second application we show in Fig.~\ref{fig:rpa} the nuclear suppression factor $R_{pA}=\der N^{pA}/N_\mathrm{coll}\der N^{pp}$ and compare with the ALICE data~\cite{ALICE:2012mj}. Even tough we have chosen to use very conservative error sets, the effect on $R_{pA}$ is very small. This suggest that $R_{pA}$ is not sensitive to the details of the dipole amplitude, and thus $R_{pA}$ is a solid CGC prediction. We have analytically shown in Ref.~\cite{Lappi:2013zma} that we get midrapidity $R_{pA}\to 1$ at all $\sqrt{s}$ at large $p_T$ which is consistent with the ALICE data.

\section*{Acknowledgements}
We thank H. Paukkunen for discussions.
This work has been supported by the Academy of Finland, projects 133005, 
267321, 273464 and by computing resources from CSC -- IT Center for 
Science in Espoo, Finland. H.M. is supported by the Graduate School of 
Particle and Nuclear Physics.

\bibliographystyle{h-physrev4mod2}
\bibliography{../../refs}

\end{document}